# Two-dimensional structures of ferroelectric domain inversion in LiNbO$_3$ by direct electron beam lithography

J. He,[a)] S. H. Tang, Y. Q. Qin, P. Dong, H. Z. Zhang, C. H. Kang, W. X. Sun, and Z. X. Shen

Department of Physics, National University of Singapore,
2 Science Drive 3, Singapore 117542, Singapore

[a)]Electronic mail: scip0128@nus.edu.sg

## ABSTRACT

We report on the fabrication of domain-reversed structures in LiNbO$_3$ by means of direct electron beam lithography at room temperature without any static bias. The LiNbO$_3$ crystals were chemically etched after the exposure of electron beam and then, the patterns of domain inversion were characterized by atomic force microscopy (AFM). In our experiment, an interesting phenomenon occurred when the electron beam wrote a one-dimensional (1-D) grating on the negative c-face: a two-dimensional (2-D) dotted array was observed on the positive c-face, which is significant for its potential to produce 2-D and three-dimensional photonic crystals. Furthermore, we also obtained 2-D ferroelectric domain inversion in the whole LiNbO$_3$ crystal by writing the 2-D square pattern on the negative c-face. Such a structure may be utilized to fabricate 2-D nonlinear photonic crystal. AFM demonstrates that a 2-D domain-reversed structure has been achieved not only on the negative c-face of the crystal, but also across the whole thickness of the crystal.

# I. INTRODUCTION

Frequency doubling of laser diodes to develop compact blue or green laser source is very attractive for applications such as high-density optical storage, printing and displays. Quasi-phase-matched second harmonic generation (QPM SHG) achieved by periodic reversal of the sign of the nonlinear coefficient $\chi^{(2)}$ of the material is a very efficient method of obtaining nonlinear frequency conversion. By appropriate choice of the period of the modulation, it is possible to phase-match any arbitrary fundamental wavelength. In the last decade, 1-D quasi-phase-matching has been well understood.[1] There are many different techniques for reversing the domain polarization: Ti-in diffusion through the +z surface near Curie temperature;[2,3] $Li_2O$-outdiffusion from the +z surface;[4,5] electric poling;[6-10] electron beam (EB) bombardment.[11-16] Recently, some works have been devoted to nonlinear frequency conversion in 2-D nonlinear photonic crystals.[17-22] Broderick's research group has fabricated a 2-D nonlinear photonic crystal: hexagonally poled lithium niobate.[18] The period of the crystal is about 18 μm and quasi-phase-matching is obtained for multiple directions of propagation. The 2-D nonlinear photonic crystal provides more freedoms and choices for multiple quasi-phase-matched frequency conversions. Though 2-D domain inversion has been obtained by electric poling technique, EB bombardment seems to be more promising due to its versatility and high resolution in particular with respect to small period.

In this article, we report on the fabrication of 1-D and 2-D ferroelectric domain-reversed structures in lithium niobate using direct EB lithography at room temperature without any static bias. The $LiNbO_3$ crystals were chemically etched after the exposure of electron beam and then the patterns of domain inversion were characterized by AFM. In our experiment, an interesting phenomenon was observed when the electron beam wrote a one-dimensional grating on the negative c-face: a two-dimensional dotted array appeared on the positive c-face. Such structures have good potential in producing 2-D and three-dimensional (3-D) nonlinear photonic crystals. In addition, we also obtained 2-D ferroelectric domain inversion in the whole LiNbO$_3$ crystal by choosing appropriate irradiation conditions. Such a structure may be utilized for designing 2-D nonlinear photonic crystals. The AFM analysis demonstrates that a 2-D domain-reversed structure has been achieved not only on the negative c-face of the crystal, but also across the whole thickness of the crystal. Such structures should be useful for various nonlinear optical applications, especially for 2-D quasi-phase-matched nonlinear frequency conversion.

## II. FABRICATION OF FERROELECTRIC DOMAIN GRATING

There are three requirements for fabricating a periodically inverted domain structure using direct electron beam lithography. Firstly, negative c-face of LiNbO$_3$ should be well-cleaned with no coating. In particular, high conductance materials, such as metals, must be completely removed from the negative c-face. Secondly, on the positive c-face of LiNbO$_3$, high conductance metals, such as Au, Al, Ag, Cr, must be coated with a tens of nanometers thick metal layer. The purpose of coating metal on the positive c-face is to homogenize electric conductance of the positive c-face of LiNbO$_3$. Furthermore, this metal on the positive c-face must be grounded. Thirdly, a periodic pattern with suitable parameters for phase matching between the interaction waves must be drawn by an electron beam on the negative c-face of LiNbO$_3$ in a vacuum. A periodically inverted domain structure is fabricated by direct drawing with an electron beam without a mask on the negative c-face. This means there is more flexibility in designing a pattern. This method is very simple and practical because it applies without voltage or heat.

In the present study, an electron beam lithography system (JEOL JBX5D2) was used for the experiment. The lithography on this system uses the vector scanning method for EB scanning and the step and repeat method for stage movement, respectively. This combination allows the entire area of the workpiece to be scanned by EB. The $LiNbO_3$ single crystals of optical grade were provided by Casix Inc. China. The thickness of the z-cut $LiNbO_3$ substrate was 0.5 mm and the +z surface was coated with an Au film of 100-nm thickness. The EB scanned area was 6 x 6 $mm^2$ with acceleration voltage 25 kV. The main irradiation parameters are given in Table I. Etching in a 1:2 mixture of hydrofluoric and nitric acids revealed the domain-reversed regions since the etching rate on the negative c-face of $LiNbO_3$ is much faster than that on the positive c-face when this enchant is used. It means that the inverted domains on the negative c-face are etched slower than the non-inverted zones while the inverted domains on the positive c-face are etched faster than the non-inverted zones. Thus, it is expectable that mountains and holes appear, corresponding to the domain-reversed regions on the negative c-face and the positive c-face, respectively. In our experiment, AFM (NanoScope IIIa Scanning Probe Microscope, Digital Instruments) is used to observe the domain patterns for confirmation.

## III. RESULTS AND DISCUSSIONS

Figs.1 (a) and (b) show, respectively, the AFM pictures of domain-inverted gratings on the negative c-face and on the positive c-face of $LiNbO_3$ crystal after 1-D area-scanning. The designed 1-D grating period is 6.5 μm and the designed grating width is 2 μm. The penetration depth of electrons into $LiNbO_3$ sample depends on the electron energy and it is estimated to be about 3 μm at an acceleration voltage of 25 kV. The domain-reversed gratings fabricated by EB exposure can be clearly seen on the negative c-face of $LiNbO_3$ crystal.

Surprisingly, the dotted grating lines are clearly seen on the positive c-face where it is 500 μm away from the EB exposing position. The separation between two adjacent hexagonal dots is about 10 μm, which is just the size of the scanning subfield. It means that domain inversion occurred discontinuously inside the crystal along the grating direction. It was found that, after etching of about 15 minutes, the height of the strips is about 0.5~1 μm on the negative c-face while the depth of the corresponding holes on the positive c-face is also about 0.5~1 μm. Hence Figs. 1 (a) and (b) demonstrate that domain inversion occurs across the whole thickness of the $LiNbO_3$ crystal by EB lithography. Fig. 2 and Fig. 3 display the comparison of etched domain-inverted patterns on +c face under different area charge density and line charge density, respectively. The domain inversion becomes more and more prominent with the increase of charge densities. When line charge density increases, the adjacent domain-inverted patterns merge with each other along the scanning direction and form line segments. Due to the continuous change of domain reversal along c direction, it is seen that this characteristic provides the possibilities to produce 3-D photonic crystals which gradually vary from 1-D written pattern on the negative c-face to 2-D pattern on the positive c-face. Fig. 4 demonstrates a possible reversed domain structure inside the crystal. The domain dimension inside the crystal is seen to be strongly dependent on the charge density used in EB lithography. Precise domain structures can be revealed by etching of cross sections along the c-axis. It indicates that adjusting of charge density could be an effective solution to control the domain dimension across the crystal thickness. Although the experimental results show that the measured grating period agrees exactly with the designed one under different charge density, lateral domain grating width is seen to be strongly related to charge density. It increases as charge density is enhanced. In some regions, adjacent grating lines merge into a big pattern (Fig. 5). We obtained similar results when we used bigger EB spot size (400 nm) and larger scanning step (150 nm). Therefore, it is difficult to obtain continuous grating lines of small period on the positive c-face by electron beam lithography.

The mechanism of ferroelectric domain inversion by EB bombardment has not been fully understood. When the EB is scanned on negative c-face of LiNbO$_3$ crystal, the electrons penetrate into the crystal for a few microns and generate an electric field across the crystal. If the induced electric field is stronger than the coercive field of the LiNbO$_3$ crystal and in the opposite direction of the spontaneous polarization, it acts as a local poling field. Domain inversion occurs beneath the negative c-face: Li ions move in the z direction into oxygen triangles and z displacements of Nb ions are also induced. Once the seed is created, the inverted domain grows toward the positive c-face and finally becomes stable. The domain inversion in segmented regions on the positive c-face seems to be related to neutralization of the deposited charges by the inversion of the spontaneous polarization since the EB scans the subfields one by one.

Since the EB system has great flexibility to scan various patterns, we attempted to achieve 2-D domain inversion directly by EB irradiation. The designed pattern is a square of 2 μm, with periodicity of 2-D structure at 6.5 μm in both x and y directions. Figs. 6 (a) and (b) illustrate, respectively, the AFM pictures of etched 2-D domain-inverted patterns on the negative and positive c-face of LiNbO$_3$ crystal after area-scanning. Arrays of hills and holes corresponding to the 2-D domain-reversed patterns are seen to appear on the negative c-face and the positive c-face, respectively. Hence Figs. 6 (a) and (b) demonstrate that 2-D periodic domain inversion occurs across the entire thickness. Our experimental results are different from recent results of Restoin's group,[23] which demonstrated a 2-D domain-inverted structure in a thin layer beneath the negative c-face. In our experiment, the designed pattern seems to induce domain-inversion with a square lattice. However, Fig. 6 (b) showed that the inverted domains on the positive c-face were hexagonally broadened, which corresponds to the hexagonal unit cell (six formula units per cell) of lithium niobate. Since LiNbO$_3$ is of the crystal class 3m and space group R3c, it shows a tendency for domain walls to form along the crystal's natural preferred domain wall orientation. Here the shape of domain-reversed pattern and symmetry of crystal are in good agreement. Moreover, in particular experimental conditions, the structure of domain-reversed patterns are shifted in adjacent grating lines indicating the movement of seed of the inverted domain grows toward the positive c-face non-vertically. The pattern size and the period were not uniform over the whole scanned surface, which is possibly due to non-uniformity of the sample, the distortion of EB position, or the fluctuation of EB current.

The irradiation parameters strongly influence the ferroelectric domain inversion process. Optimization of the charge density is an important requirement for obtaining the domain-inverted structures. Excessive charge density results in the deformation of the domain-inverted grating while insufficient charge density does not induce the domain inversion. For a fixed charge density, too big a current will cause cracks. In addition, the scanning velocity $v_s$ is directly related to the charge density $Q_s$ and the electron beam current I:

$$v_s = I/Q_s \qquad (1)$$

where $v_s$ is area-scanning velocity (line-scanning velocity) and correspondingly $Q_s$ is area charge density (line charge density). A larger current will correspond to faster scanning velocity. Hence all parameters must be adjusted together to obtain optimized ferroelectric domain inversion.

## IV. CONCLUSIONS

In summary, 2-D periodic domain-inverted structures have been fabricated by direct electron beam lithography at room temperature without applying a voltage. The irradiation parameters to realize domain inversion were determined. Study of structures by AFM confirmed that the domain inversion occurred across the whole thickness of the crystal. This point is especially important to achieve good conversion efficiencies for various QPM nonlinear frequency conversions. And we expect that such structures could be used for a broad range of 2-D as well as potential 3-D nonlinear photonic crystals. Work is in progress to improve the uniformity of the periodical domain inversion and to enhance the efficiency of nonlinear frequency conversion.

## ACKNOWLEDGMENTS

This work was supported by the Research Grant R-144-000-027-112 and R-144-000-074-422 from National University of Singapore.

**REFFERENCES**

1. M. M. Fejer, G. A. Magel, D. H. Jundt, and R. L. Byer, IEEE J. Quantum Electron. **28**, 2631 (1992).

2. S. Miyazawa, J. Appl. Phys. **50**, 4599 (1979).

3. S. Thaniyavarn, T. Finddakly, D. Boeher, and J. Moen, Appl. Phys. Lett. **46**, 933 (1985).

4. K. Yamamoto, K. Mizunchi, K. Takeshige, Y. Sasai, and T. Tanichi, J. Appl. Phys. **70**, 1947 (1991).

5. C. J. Van Der Poel, J. D. Bierlen, J. B. Brown, and S. Colak, Appl. Phys. Lett. **57**, 2074 (1992).

6. L. E. Myers, R. C. Eckardt, M. M. Fejer, R. L. Byer, W. R. Bosenberg, and J. W. Pierce, J. Opt. Soc. Am. B **12**, 2102 (1995).

7. G. D. Miller, R. G. Batchko, M. M. Fejer, and R. L. Byer, Proc. SPIE **2700**, 34 (1996).

8. S.-n. Zhu, Y.-y Zhu, Yi.-q Qin, H.-f Wang, C.-z Ge, and N.-b Ming, Phys. Rev. Lett. **78**, 2752 (1997).

9. Shi-ning Zhu, Yong-yuan Zhu, and Nai-ben Ming, Science **278**, 843 (1997).

10. G. Rosenman, K. Garb, A. Skliar, M, Oron, D. Eger, and M. Katz, Appl. Phys. Lett. **73**, 865 (1998).

11. C. Restoin, C. Darraud-Taupiac, J. L. Decossas, and J. C. Vareille, J. Appl. Phys. **88**, 6665 (2000).

12. Wei-yung Hsu and Mool C. Gupta, Appl. Phys. Lett. **60**, 1 (1992).

13. Mool C. Gupta, W. P. Risk, Alan C. G. Nutt and S. D. Lau, Appl. Phys. Lett. **63**, 9 (1993).

14. M. Fujimura and T. Suhara, J. Lightwave Technol. **11**, 1360 (1993).

15. S. Kurimura, I. Shimoya and Y. Uesu, Jpn. J. Appl. Phys. **35**, L31, Part 2, No. 1A, (1996).

16. A. Nakao, K. Fujita, T. Suhara, and H. Nishihara, Jpn. J. Appl. Phys. **37**, 845, Part 1, No. 3A, (1998).

17. V. Berger, Phys. Rev. Lett. **81**, 4136 (1998).

18. N. G. R. Broderick, G. W. Ross, H. L. Offerhaus, D. J. Richardson and D. C. Hanna, Phys. Rev. Lett. **84**, 4345 (2000).

19. Aref Chowdhury, Susan C. Hagness, and Leon McCaughan, Opt. Lett. **25**, 832 (2000).

20. A. Chowdhury, C. Staus, B. F. Boland, T. F. Kuech and L. McCaughan, Opt. Lett. **26**, 1353 (2001).

21. Solomon S. Saltiel and Yuri S. Kivshar, Opt. Lett. **25**, 1204 (2000).

22. Martijn de Sterke, Solomon S. Saltiel and Yuri S. Kivshar, Opt. Lett. **26**, 539 (2001).

23. C. Restoin, S. Massy, C. Darraud-Taupiac, and A. Barthelemy, Opt. Mater. **22**, 193 (2003).

Figure Captions:

Fig. 1 AFM pictures of etched domain-inverted gratings fabricated by 1-D EB area scanning (area charge density: 50 μC/cm$^2$, current: 500pA, beam diameter: 70nm); (a) Z- surface, (b) Z+ surface. The designed and measured grating period is 6.5 μm.

Fig. 2 Optical micrographs of etched domain-inverted patterns on Z+ surface under different area charge density (current: 500 pA, beam diameter: 70 nm); (a) 70, (b) 150, (c) 250, and (d) 350 μC/cm$^2$. The designed and measured grating period is 6.5 μm.

Fig. 3 Optical micrographs of etched domain-inverted patterns on Z+ surface under different line charge density (current: 500 pA, beam diameter: 70 nm); (a) 30, (b) 70, (c) 110, and (d) 170 nC/cm. The designed and measured grating period is 6.5 μm.

Fig. 4 Schematic representation of reversed domain structures inside LiNbO$_3$ crystal by EB irradiation.

Fig. 5 Optical micrograph (a) and (b) are etched domain-inverted patterns in different regions on Z+ surface fabricated by EB line-scanning (line charge density: 170 nC/cm, current: 500 pA, beam diameter: 70 nm). The designed grating period is 6.5 μm.

Fig. 6 AFM pictures of etched 2-D periodic domain-inverted structure fabricated by 2-D EB area scanning (area charge density: 500 μC/cm$^2$, current: 500 pA, beam diameter: 400 nm); (a) Z- surface, (b) Z+ surface. The designed and measured 2-D lattice period is 6.5 μm.

TABLE I. Main EB parameters used to achieve ferroelectric domain inversion.

|  | Area scanning mode | Line scanning mode |
|---|---|---|
| Accelerating voltage (kV) | 25 | 25 |
| EB current (pA) | 500 | 500 |
| Charge density | 50~500 $\mu C/cm^2$ | 30~170 nC/cm |
| Spot size (nm) | 70/400 | 70/400 |
| Scanning step (nm) | 15/150 | 15/150 |
| Grating period ($\mu$m) | 6.5 | 6.5 |

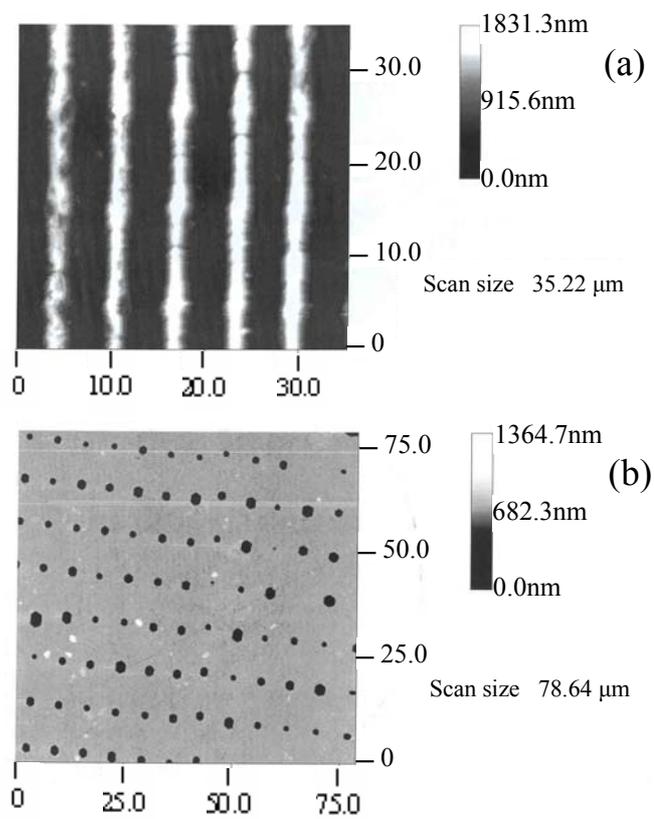

Fig. 1

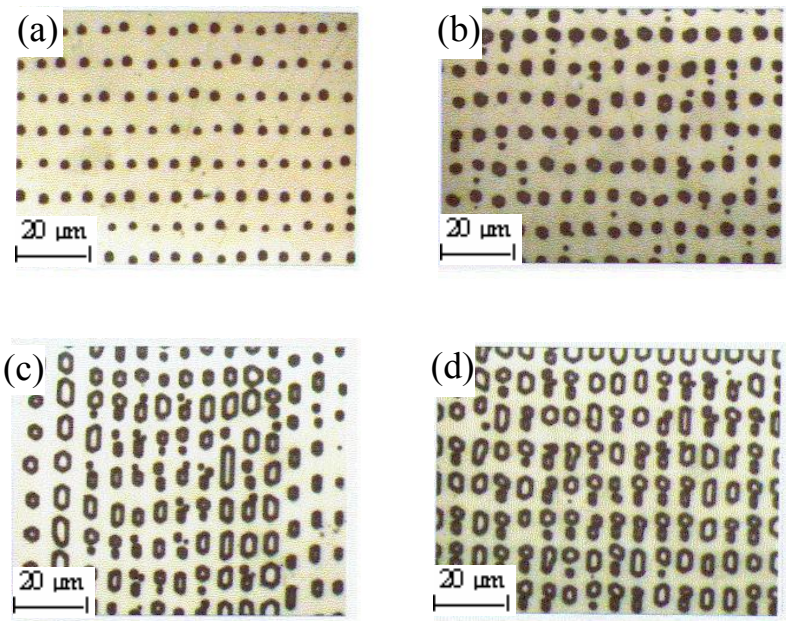

Fig. 2

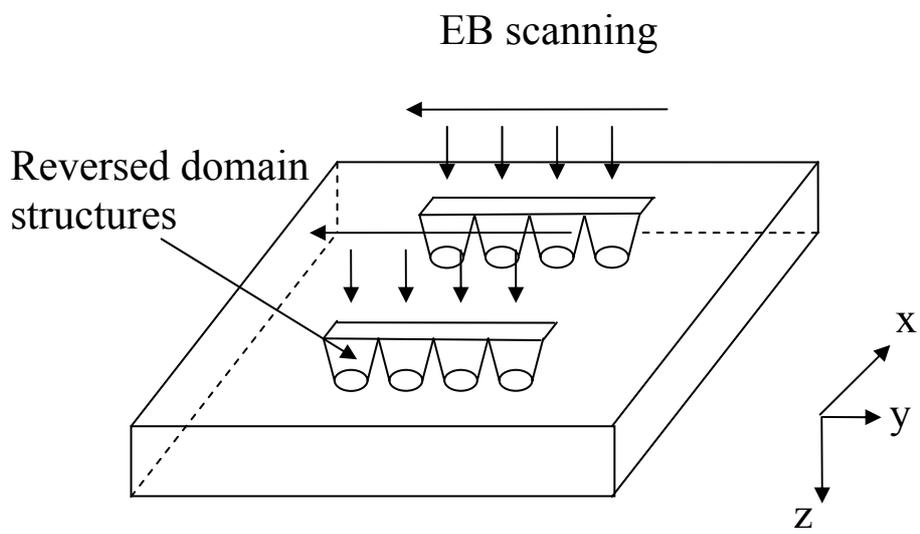

Fig. 4

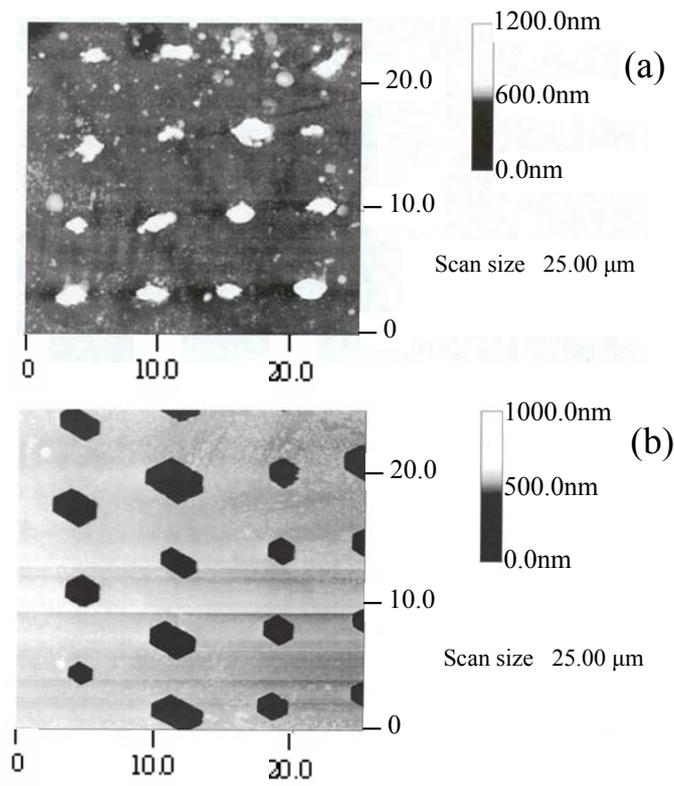

Fig. 6